\documentclass[12pt]{article}
\usepackage{amsmath,amssymb,amscd,latexsym,euscript,mathrsfs}
\usepackage[all]{xy}

\usepackage[cp1251]{inputenc}

\newcommand{\Set}{{\rm Set}}

\newcommand{\Tran}{{\rm Tran}}

\newcommand{\NN}{{\,\mathbb N}}

\newcommand{\mN}{{\,\cal N}}

\newcommand{\ZZ}{{\,\mathbb Z}}
\newcommand{\mA}{{\mathcal A}}

\newcommand{\fM}{{\mathfrak{M}}}

\newtheorem{theorem}{\bf Theorem}[section]
\newtheorem{lemma}[theorem]{\bf Lemma}
\newtheorem{proposition}[theorem]{\bf Proposition}
\newtheorem{corollary}[theorem]{\bf Corollary}
\newtheorem{definition}{\sc Definition}[section]
\newtheorem{example}[definition]{\sc Example}
\newtheorem{remark}[definition]{\sc Remark}

\def\leq{\leqslant}
\def\geq{\geqslant}

\begin{document}

\title{Homology and Bisimulation of Asynchronous Transition Systems and Petri Nets
\footnote{This work was performed as a part of the Strategic Development Program at the National
Educational Institutions of the Higher Education, 
N 2011-PR-054}}

\author{Ahmet A. Husainov}

\maketitle

\begin{abstract}
Homology groups of labelled asynchronous transition systems and Petri nets are introduced. 
Examples  of computing the homology groups are given.
It is proved that if labelled asynchronous transition systems are bisimulation equivalent, 
then  they have isomorphic homology groups.
A method of constructing a Petri net with given homology groups is found.
\end{abstract}

2000 Mathematics Subject Classification 18G35, 18B20, 55U10, 55U15, 68Q85


\section*{Introduction}

The paper is devoted to the application of algebraic topology methods 
for classification and studying the mathematical
models of concurrency.
We consider
asynchronous transition systems with label functions on events.
Our purpose is to construct a homology theory of labelled asynchronous transition  systems
for which any bisimulation equivalent asynchronous transition  systems have
isomorphic homology groups.

We consider a categorical notion of the bisimulation defined by open maps \cite{joy1994}.
It was proved in \cite{joy1994},
that in the case of labelled transition systems this definition 
coincides with a strong bisimulation of R. Milner \cite{mil1989}.
A characterization of the bisimilation equivalence for asynchronous transition  systems
was given in \cite{nie1996}.

Homology groups have no less than important for the classification and studying 
the properties of concurrent
systems. 
 In particular, they have been applied
in the work \cite {her1999}
to characterize the condition of solvability for some classes of problems in parallel
distribution systems.

In \cite{gou1992}, E. Goubault and T. P. Jensen applied homology 
groups for studying higher dimensional automata.
There  were obtained
some signs of bisimulation equivalence for the higher dimensional automata 
in terms of the homology groups \cite[Prop. 10]{gou1992}. The results were developed in the
\cite {gou1995}. In a survey \cite{faj2006}, open questions were marked on
the relationship of the Goubault homology \cite{gou1995} with directed homotopy.
The Goubault homology
have been applied also to prove of homotopy properties
for higher dimensional automata in the \cite{gou2009}.
Communications between homotopy and bisimilarity
of higher dimensional automata was researched in \cite{fah2012}.

These groups were used to find signs of parallelizable asynchronous systems in \cite{X20082} and
 were regarded as the homology groups of a topological space of
intermediate states for an asynchronous system in \cite{X20121}. 
An algorithm for computing the homology groups was developed in \cite{X2012}.

In this paper, we study the homology of the {\em labelled} asynchronous transition  systems and Petri nets.

We work in the category of asynchronous transition systems considered in \cite{win1995}.
But we call them simply {\em asynchronous systems}. 
Note that M.A. Bednarczyk \cite{bed1988} studied the broader category of asynchronous systems.
Using results of M. Nielsen and G. Winskel \cite{nie1996}, 
we study open morphisms.
We introduce homology groups for labelled asynchronous transition  systems and Petri nets.
We prove that $Pom_L$-bisimilar asynchronous transition  systems have isomorphic homology groups
(Theorem \ref{isohomol} and Corollary \ref{isoall}).
We give some examples of computing the homology groups of asynchronous transition  systems and Petri nets.
We prove that for an arbitrary finite sequence of finitely generated Abelian groups
$A_0$, $A_1$, $A_2$,  \ldots 
where $A_0$ is free and  not equal $0$ there exists a labelled Petri net the
$i$th homology groups of which are isomorphic to $A_i$ for all  $i\geq 0$.

\tableofcontents

\section{Asynchronous  systems and trace monoid actions}

Let us recall some facts on the mathematical models of concurrency 
\cite{nie1996}, \cite {win1995}, \cite{bed1988}. We study asynchronous
systems as trace monoids with partial action on sets.

\subsection{State spaces and asynchronous systems}

\begin{definition}
{\em A state space} $(S,E,I,\Tran)$ consists of a set $S$ of {\em states}, 
a set $E$ of {\em events} with a symmetric irreflexive relation $I\subseteq E\times E$ 
of {\em independence}, and a {\em transition relation} $\Tran\subseteq S\times E\times S$.
 The following axioms must be satisfied:
\begin{enumerate}
\item  If $(s,a,s')\in \Tran$ $\&$ $(s,a,s'')\in \Tran$, then $s'=s''$.\label{prop1}
\item  If $(a,b)\in I ~ \& ~(s, a, s')\in \Tran ~\& ~
(s', b, s'')\in \Tran$, then there exists $s_1\in S$ such that $(s,b,s_1)\in \Tran$
$\&$ $(s_1, a, s'')\in \Tran$. (See Fig. \ref{axiom2}) \label{prop2}
\end{enumerate}
\end{definition}
\begin{figure}[ht]
$$
\xymatrix{
& s'\ar[rd]^b\\
s \ar[ru]^a \ar@{-->}[rd]_b && s''\\
& s_1 \ar@{-->}[ru]_a
}
$$
\caption{To Axiom {(\ref{prop2})}}.
\label{axiom2}
\end{figure}

Triples $(s, e, s')\in \Tran$ are denoted by $s\stackrel{e}\to s'$ and called {\em transitions} .

\begin{definition}
{\em Asynchronous system} $\mA= (S, s_0, E, I, \Tran)$ is a state space
$(S, E, I, \Tran)$ with a distinguished {\em initial state}
 $s_0\in S$. Moreover, 
for every $a\in E$, there must be $s_1, s_2\in S$ satisfying $(s_1, a, s_2)\in Tran$.
\end{definition}

\begin{definition}
A {\em morphism} between state spaces 
$$
(\sigma, \eta):(S,E,I,\Tran)\to (S',E',I',\Tran')
$$ 
is a pair consisting of a partial map $\eta: E\rightharpoonup E'$
and a map $\sigma: S\to S'$ satisfying the following conditions
\begin{enumerate}
\item  for any triple  $(s_1, e, s_2)\in \Tran$,
there is the following alternative
$$
\left\{
\begin{array}{cl}
(\sigma(s_1), \eta(e), \sigma(s_2))\in \Tran', & \mbox{ if the value } \eta(e) \mbox{ is defined},\\
\sigma(s_1)=\sigma(s_2), & \mbox{ if } \eta(e) \mbox{ is not defined};
\end{array}
\right.
$$
\item for all $(e_1,e_2)\in I$, if $\eta(e_1)$ and $\eta(e_2)$ both are defined,
then
$(\eta(e_1),\eta(e_2))\in I'$.
\end{enumerate}
\end{definition}

Let $\mA=(S,s_0, E, I, \Tran)$ and $\mA'=(S',s'_0, E', I', \Tran')$ be asynchronous systems.
A morphism of asynchronous systems  $(\sigma, \eta): \mA\to \mA'$ is a mophism 
 $(\sigma, \eta): (S, E, I, \Tran)\to (S', E', I', \Tran')$ 
between the state spaces such that  $\sigma(s_0)=s'_0$.

\subsection{Asynchronous systems and partial actions of trace monoids}

Below, throughout the paper, we will denote $\mA=(S,s_0, E, I, \Tran)$ and $\mA'=(S',s'_0, E', I', \Tran')$.

For an arbitrary category $\cal C$, let ${\cal C}^{op}$ be the opposite category.  

Denote by $PSet$ the category of sets and partial maps.
Let $M$ be a monoid considered as the category
with a single object.
A {\em partial right action of a monoid $M$ on a set $S$} is
a functor $M^{op} \to PSet$, the value of which on the single object is equal to $S$.
The functor assigns to each morphism $\mu \in M$ a partial map $S \rightharpoonup S$
the values of which defined on $s \in S$ are denoted by $s \cdot \mu$.
The category $PSet$ is equivalent to the category
of pointed sets and pointed maps \cite{win1995}. If we leave
pointed sets, whose distinguished points  are equal to a fixed common point $*$,
then we obtain a category isomorphic to the category $PSet$. We denote this category by $\Set_*$.
The isomorphism allows us to consider a partial right action of $ M $ on $ S $ as a
functor $ M^{op} \to \Set_* $. We denote this functor by $ (M, S_*) $.
For each $\mu\in M$, its value $(M,S_*)(\mu)$ is the map denoted by $s \mapsto s\cdot\mu$
for all $s\in S_*$.

In particular, the state space can be considered
as a set with a partial action of a {\em trace monoid}.
Let us recall the definition of a trace monoid \cite{die1997}.

Let $E$ be a set with a symmetric irreflexive relation $I \subseteq E \times E$.
Denote by $E^*$ a free monoid of words with the letters of $E$. 
 Elements $ a, b \in E $
are  {\em independent} if $ (a, b) \in I $. We define an equivalence relation on $ E^* $
assuming $ w_1 \equiv w_2 $ if the word $ w_2 $ can be obtained from $ w_1 $ by a finite sequence
permutations of adjacent independent elements.
Let $ [w] $ be the equivalence class of $ w \in E^* $.
It is easy to see that the operation $ [w_1] [w_2] = [w_1 w_2] $ transforms
the set of equivalence classes $ E^* / \equiv $ in a monoid. This monoid is
called a trace monoid $ M (E, I) $.

Let $(S, E, I, \Tran)$ be a state space.
For any $ s \in S $ and $ e \in E $, there exists at most one $ s' \in S $
for which $ (s, e, s') \in \Tran $. In this case, we set $ s \cdot e = e '$.
If $\Tran$ does not contain such a triple, then let $ s \cdot e = * $.
Now we can assign to each state space $ (S, E, I, \Tran) $
the partial action
$ (M (E, I), S_*) $ defined as 
$ (s, [e_1 \cdots e_n]) \mapsto (\ldots ((s \cdot e_1) \cdot e_2) \ldots \cdot e_n) $.
Any asynchronous system can be considered as a partial action $(M(E,I), S_*)$ of the trace monoid on $S$
with initial element $s_0\in S$.
It follows from the definition of action that the formula $s\cdot e\in S$ is equivalent to 
$(\exists t\in S) (s, e, t)\in \Tran$. 
This formula means that
the value $ s \cdot e $ is defined, but  $ s \cdot e = * $ means
that this value is not defined.
The morphism between asynchronous systems  $\mA\to \mA'$ can be defined as a pair of maps 
$\sigma: S\to S'$, $\eta: E\to E'\cup \{1\}$ for which
\begin{itemize}
\item the map $ \eta $ can be extended to a homomorphism of monoids $ M (E, I) \to M (E ', I') $;
\item for every $ s \in S $ and $ e \in E $ satisfying $ s \cdot e \in S $, it is true that
$ \sigma (s) \cdot \eta (e) \in S ~ \& ~ \sigma (s) \cdot \eta (e) = \sigma (s \cdot e) $;
\item $\sigma(s_0)= s'_0$.
\end{itemize}

\subsection{Open morphisms}

A state $s\in S$ of asynchronous system $\mA$ is {\em reachable} if there exists a finite sequence of transitions
$s_0 \stackrel{e_1}\to s_1 \stackrel{e_2}\to s_2 \to 
\cdots \to s_{n-1}\stackrel{e_n}\to s$.

If we want to emphasize that the map $ f: X \to Y $ is defined on all elements of $X$, then we call
it {\em total}.

\begin{definition}\label{defopen}
A morphism of asynchronous systems $(\sigma, \eta): \mA\to \mA'$ is {\em open},
if it has the following properties:
\begin{enumerate}
\item \label{property1} $\eta: E\to E'$ is total;
\item \label{property2} for all a state $s\in S$ and transition $(\sigma(s), e', u')\in \Tran'$, there exists
 $(s,e,u)\in \Tran$ for which $\eta(e)=e'$ and $\sigma(u)=u'$;
\item \label{property3} for any reachable $s\in S$, if $(s,e_1,u)\in \Tran$ and 
 $(u,e_2,v)\in \Tran$ and 
$(\eta(e_1), \eta(e_2))\in I'$, then $(e_1,e_2)\in I$. 
\end{enumerate}
\end{definition}

The property (\ref{property2}) can be shown visually by drawing
$$
\xymatrix{
s \ar@{-->}[dd]_{\exists\,e} & & & \sigma(s)\ar[dd]^{\forall\,e'}\\
  & \ar@{|->}[r]^{\eta} &\\
u 		& & &	u'
}
$$

For any asynchronous system $\mA=(S,s_0, E,I, \Tran)$ and a reachable $s\in S$, 
we let $\mA(s)= (S,s, E,I, \Tran)$.
In particular, $\mA(s_0)=\mA$. 

\begin{proposition}\label{opensub}
For any open morphism $(\sigma, \eta): \mA\to \mA'$  of asynch\-ro\-no\-us systems and  
 a reachable state $s\in S$,
the morphism $(\sigma, \eta): \mA(s)\to \mA'(\sigma(s))$ is open.
\end{proposition}

\section{Bisimulation equivalence of labelled asynchronous systems}

In this section, we consider $Pom_L$-bisimilar labelled asynchronous systems.

\subsection{Labelled asynchronous systems}

A {\em labelled asynchronous system}  $(\mA, \lambda, L)$ consists of  
 an asynchronous system $\mA$ with an arbitrary set $L$ of {\em labels} and a map
 $\lambda: E\to L$ called {\em label function}.
Each asynchronous system can be considered as labelled where the set $L=pt$ consists of a single label.
In this sense, according to \cite [Prop. 16]{nie1996}, 
open morphisms are precisely $Pom_{pt}$-open morphisms.

Let $(\mA, \lambda, L)$ and $(\mA', \lambda', L)$  be labelled asynchronous systems.
 A morphism  $(\sigma, \eta): \mA\to \mA'$
   {\em preserves labels }, if for all $e\in E$, it satisfies to equality
$\lambda(e)= \lambda'(\eta(e))$. In this case, the pair $(\sigma, \eta)$ is called a 
{\em morphism of labelled asynchronous systems}
$(\mA, \lambda, L) \to (\mA', \lambda', L)$.

The following statement is a reformulation of the characterization of $ Pom_L $-morphisms
given in \cite[Prop.16]{nie1996}.

\begin{proposition}
A morphism $(\sigma, \eta): (\mA,\lambda, L ) \to (\mA',\lambda', L)$ 
between labelled asynchronous systems is $Pom_L$-open 
if and only if the morphism $(\sigma, \eta): \mA \to \mA'$ is open and preserves labels. 
\end{proposition}

This proposition allows us to mean by $Pom_L$-open morphisms the open morphisms, preserving labels.

\begin{definition}\cite{nie1996}
Let $(\mA, \lambda, L)$ and $(\mA', \lambda', L)$ be labelled asynchronous systems.
If there exists a labelled asynchronous system  
$(\mA'', \lambda'', L)$ with $Pom_L$-open morphisms  
$(\mA'', \lambda'', L) \stackrel{(\sigma,\eta)}\to (\mA, \lambda, L)$ and 
$(\mA'', \lambda'', L) \stackrel{(\sigma',\eta')}\to (\mA', \lambda', L)$, 
then  $(\mA, \lambda, L)$ and $(\mA', \lambda', L)$ are called {\em $Pom_L$-bisimilar}.
\end{definition}

\begin{proposition}\label{bissub}
Let $(\mA, \lambda, L)$ and $(\mA', \lambda', L)$ be $Pom_L$-bisimilar labelled asynchronous systems.  
For every $a_1\in E$ satisfying $s_0\cdot a_1\in S$, there exists $a'_1\in E'$ such that  
the following two properties hold: 
\begin{itemize}
\item $s'_0\cdot a'_1\in S'$;
\item labelled asynchronous systems $(\mA(s_1), \lambda, L)$
 and $(\mA(s'_0\cdot a'_1), \lambda', L)$ are $Pom_L$-bisimilar.
\end{itemize}
\end{proposition}
{\sc Proof.} Given labelled asynchronous systems are 
$Pom_L$-bisimilar. Hence, there are 
$(\mA'', \lambda'', L)$ and $Pom_L$-open morphisms
$$
	(\mA, \lambda, L) \stackrel{(\sigma,\eta)}\longleftarrow 
	(\mA'', \lambda'', L) \stackrel{(\sigma',\eta')}\longrightarrow 
(\mA', \lambda', L). 
$$
Morphism $(\sigma,\eta)$ is open. It follows by property  (\ref{property2}) 
of Definition \ref{defopen} that there exists a transition $(s''_0, a''_1, s''_1)$ 
satisfying conditions
$\eta(a''_1)=a_1$ and $\sigma(s''_1)=s_1$ (Fig. \ref{pic15}). In other words, 
there exists $a''_1\in E''$ such that $\eta(a''_1)=a_1$ and $\sigma(s''_0\cdot a''_1)= s''_1$.
By Proposition \ref{opensub}, 
the morphism $(\sigma,\eta): \mA''(s''_1)\to \mA(s_1)$ is open.
\begin{figure}[ht]
$$
\xymatrix{
s_0 \ar[d]_{a_1} & s''_0 \ar@{|->}[l]_{\sigma} \ar@{-->}[d]^{a''_1} 
			\ar@{|->}[r]^{\sigma'} & s'_0\ar@{-->}[d]^{\eta'(a''_1)}\\
s_1 & s''_1 \ar@{|->}[l]^{~~\sigma} \ar@{|->}[r]_{\sigma'} & \sigma'(s''_1)
}
$$
\caption{To the construction of open morphisms.}\label{pic15}
\end{figure}
The map $\sigma'$ of the morphism $(\sigma',\eta'): \mA'' \to \mA'$ is total. 
It follows that $\sigma'(s''_1)\in S'$. By Proposition \ref{opensub}, the morphism
 $(\sigma,\eta): \mA''(s''_1)\stackrel{(\sigma',\eta')}\to \mA'(\sigma'(s''_1))$ is open.
The morphisms $(\sigma,\eta)$ and $(\sigma',\eta')$ preserve labels. 
By putting $a'_1= \eta'(a''_1)$ and $s'_1= \sigma'(s''_1)$, we obtain the desired.
\hfill $\Box$

\begin{corollary}\label{bisall}
Let $(\mA, \lambda, L)$ and $(\mA', \lambda', L)$ be labelled asynchronous systems.
For every $w= a_1\cdots a_k\in E^*$ with $k\geqslant 0$ satisfying the condition 
$s_0\cdot w\in S$, there exists a word $w'= a'_1\cdots a'_k\in E'^*$ such that the 
following two properies hold:
\begin{itemize} 
\item $s'_0\cdot w'\in S'$;
\item the labelled asynchronous systems 
$(\mA(s_0\cdot w), \lambda, L)$ and $(\mA'(s'_0\cdot w'), \lambda', L)$ are $Pom_L$-bisimilar.
\end{itemize}
\end{corollary}
{\sc Proof.} 
For $k=0$, the word $w$ is empty, that is $w=1$. Taking $w'=1$, we get the $Pom_L$-bisimilar 
labelled asynchronous systems $(\mA, \lambda, L)$ and $(\mA', \lambda', L)$. 
For $k=1$, the assertion follows from Proposition \ref{bissub}.
Assuming that the assertion is true for some $k>0$, 
we can prove by Proposition \ref{bissub}, that it holds for $k+1$.
So, it is true for all $ k \geq 0 $.
\hfill $\Box$

\subsection{Open maps and surjectivity}

Let $\mA$ be an asynchronous system.
Denote by
$Q_0(\mA)= S(s_0)$ the set of all reachable states $s\in S$.
For every 
 $n> 0$, we consider sets 
\begin{multline*}
Q_n(\mA)= \{(s, e_1, \cdots, e_n)\in S(s_0)\times E^n  
\\
s\cdot e_1\cdots e_n \in S~ \& ~(e_i, e_j)\in I \mbox{ for all }
1\leq i< j\leq n \}
\end{multline*}

Let $(\sigma,\eta): \mA\to \mA'$ be a morphism of asynchronous system.
If $\eta: E\to E'$ is total, 
 then for all $n\geq 0$ 
the maps 
$Q_n(\sigma, \eta): Q_n(\mA)\to Q_n(\mA')$ are defined by the formula
$$
Q_n(\sigma, \eta)(s, e_1, \cdots, e_n) = (\sigma(s), \eta(e_1), \cdots, \eta(e_n)).
$$
\begin{lemma}
If $(\sigma,\eta): \mA\to \mA'$ is open, then for every reachable 
$s'\in S'$ there exists $s\in S$ such that $\sigma(s)=s'$.
\end{lemma}
{\sc Proof.} 
We have $\sigma(s_0)=s'_0$. 
If $s'$ is reachable, then there exists a path 
$\sigma(s_0)= s'_0 \stackrel{a'_1}\to s'_1 \to \cdots \to s'_{n-1} \stackrel{a'_n}\to s'_n$.
The morphism $(\sigma, \eta)$ is open. Hence for  $a'_1$ and $s'_1$, 
there are $a_1$ and $s_1$ satisfying 
$\eta(a_1)=a'_1$ and $\sigma(s_1)=s'_1$. Then we find $a_2\in E$ satisfying  $\eta(a_2)=a'_2$.
And so on till we find $a_n\in E$ such that $\eta(a_n)=a'_n$ and $\sigma(s_n)=s'$.
Desired element $s$ will be equal to $s_n$.
\hfill $\Box$

\begin{proposition}\label{opensur}
If a morphism $(\sigma,\eta): \mA\to \mA'$ is open, then
the maps  $Q_n(\mA)\to Q_n(\mA')$
are surjective.
\end{proposition}
{\sc Proof.} Prove for  $n=0$. If $s'$ is reachable, then there exists a path 
$$
	\sigma(s_0)=s'_0 \stackrel{a'_1}\to s'_1 \stackrel{a'_2}\to \ldots 
\stackrel{a'_n}\to s'_k =s'.  
$$
There are $a_1\in E$ and $s_1\in S$ for which  $\eta(a_1)=a'_1$
and $(\sigma,\eta)(s_0 \stackrel{a_1}\to s_1)= (s'_0 \stackrel{a'_1}\to s'_1)$:
$$
\xymatrix{
s_0 \ar[d]^{a_1} \ar@{|->}[r]^{\sigma} & s'_0\ar[d]^{a'_1}\\
s_1 \ar@{|->}[r]^{\sigma} & s'_1
}
$$
We have $\sigma(s_1)=s'_1$.
There are $a_2\in E$ and $s_2\in S$ satisfying $\sigma(s_2)=s'_2$
 and $\eta(a_2)=a'_2$ and so on.
By induction, we obtain  $s_k\in S$ such that $\sigma(s_k)= s'_k=s'$.
Therefore, $\sigma: S(s_0)\to S'(s'_0)$ is surjective.

For $n=1$, the map 
$\{(s,e_1)| s e_1 \in S\}\to\{ (\sigma(s),e'_1)| \sigma(s) e'_1 \in S'\}$ 
is surjective by property (\ref{property2}) of open morphisms.

Let $n\geq 2$. 
For each  $s\in S(s_0)$, consider the set
\begin{multline*}
Q_n(\mA, s)= \{(s, e_1, \cdots, e_n)\in \{s\}\times E^n ~| \\
s\cdot e_1\cdots e_n \in S ~\&~ (e_i, e_j)\in I \mbox{ for all }
1\leq i< j\leq n 
\}
\end{multline*}
and
\begin{multline*}
Q_n(\mA', \sigma(s)) =
\{
(\sigma(s), e'_1, \cdots, e'_n)\in \{\sigma(s)\}\times {E'}^n |\\
 \sigma(s)\cdot e'_1\cdots e'_n \in S' ~ \& 
~(e'_i, e'_j)\in I \mbox{ for all }
1\leq i< j\leq n.
\}
\end{multline*}
For any 
$(\sigma(s), e'_1, \cdots, e'_n)\in Q_n(\mA', \sigma(s))$,
there are  $e_1$, $e_2$, \ldots, $e_n \in E$ for which
 $s_1= s\cdot e_1\in S$, $s_2= s\cdot e_1 e_2\in S$, \ldots, 
$s_n = s\cdot e_1\cdots e_n\in S$, 
wherein $\eta(e_1)= e'_1$, \ldots, $\eta(e_n)= e'_n$.

By induction on $n$, we will prove 
that $(e_i, e_j)\in I$ for all $1\leq i< j\leq n$.
For this purpose, we assume that $(e_i, e_j)\in I$ for all $1\leq i< j\leq n-1$.
And we show that $(e_i, e_n)\in I$ for all $1\leq i\leq n-1$.
We have $(s_{n-2}, e_{n-1}, s_{n-1})\in \Tran$, $(s_{n-1}, e_n, s_n)\in \Tran$, and 
$(\eta(e_{n-1}), \eta(e_n))\in I'$. It follows by the property (\ref{property3}) 
 that $(e_{n-1}, e_n)\in I$. By Axiom (\ref{prop2}) for a state space, there is 
 $t\in S$ such that $(s_{n-2}, e_n, t)\in \Tran$ 
 and $(t, e_{n-1}, s_n)\in \Tran$. It follows from $(\eta(e_{n-2}), \eta(e_n))\in I'$,
that $(e_{n-2}, e_n)\in I$. Again by Axiom (\ref{prop2}), there is 
 $t_1\in S$ such that $(s_{n-3}, e_n, t_1)\in \Tran$
 and $(t_1, e_{n-2}, s_n)\in \Tran$. It follows from  $(\eta(e_{n-3}), \eta(e_n))\in I'$,
that $(e_{n-3}, e_n)\in I$, and so on. In the end, we obtain $(e_i, e_n)\in I$
for all $1\leq i \leq n-1$. Consequently 
$(e_i, e_j)\in I$ for all $1 \leq i<j \leq n$.
Thus, $(s, e_1, \ldots, e_n)\in Q_n(\mA, s)$.
Therefore for every $(s', e'_1, \ldots, e'_n)\in Q_n(\mA)$,
there is $(s, e_1, \ldots, e_n)\in Q_n(\mA)$ mapped to  
$(s', e'_1, \ldots, e'_n)\in Q_n(\mA)$. 
\hfill $\Box$

\begin{remark}
The converse is not true. 
There are morphisms $ (\sigma, \eta) $,
for which the map $ Q_n (\sigma, \eta) $ is surjective for all $ n \geq 0 $,
but the $ (\sigma, \eta) $ is not $ Pom_ {pt}$-open.
For example, $S=\{s_0\}$, $E=\{a,b,c\}$, $I=\{(a,b), (b,a)\}$,
$S'=\{s'_0\}$,  $E'=\{a',b'\}$, $I'=\{(a',b'), (b',a')\}$. 
Figure \ref{actgrin} shows the independence graphs  and the 
map $ \eta: E \to E'$.
\begin{figure}[ht]
$$
 \xymatrix{
b\ar@{-}[d] &   & \ar@{--}[r]&\ar@{-->}[r] &  & \eta(b) \ar@{-}[d]\\
a           & c & \ar@{--}[r]&\ar@{-->}[r] &  & \eta(a)=\eta(c)
}
$$
\caption{Example of surjection which is not $Pom_{pt}$}\label{actgrin}
\end{figure}
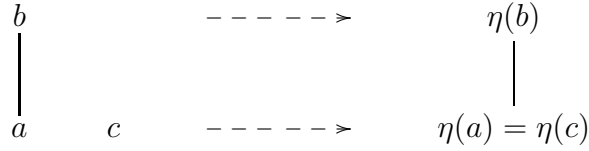
We have $ (\eta (b), \eta (c)) \in I'$, but $ (b, c) \notin I $.
Hence, the morphism $ (\sigma, \eta) $ is not open.
\end{remark}

For an reachable state $s\in S$ of asynchronous system $\mA=(S,s, E, I, \Tran)$, let 
$\mA(s)=(S,s, E, I, \Tran)$ be the asynchronous system which differs 
only by the initial state.


\begin{corollary}
If $(\sigma, \eta): \mA \to \mA'$ is open, then for each reachable state $s\in S$, 
the maps  $Q_n(\mA(s))\to Q_n(\mA'(\sigma(s)))$ are surjective for all $n\geq 0$.
\end{corollary}

\section{Homology groups of asynchronous systems}

We introduce the homology groups of labelled asynchronous systems.
We will prove that bisimulation equivalence is stronger than property
to have isomorphic homology groups.

\subsection{Computing homology groups of simplicial schemes}

Recall that a {\em simplicial scheme} $ (A, \fM) $ consists of a set $A$ of {\em vertices} 
and a set $ \fM $ of finite nonempty subsets $ S \subseteq A $
satisfying the following conditions
\begin{itemize}
\item $(\forall a\in A)$  $\{a\}\in \fM$,
\item $(\forall S, S'\subseteq A)~ S\in \fM ~\&~ S'\subseteq S 
\Rightarrow S'\in \fM$.
\end{itemize}
The elements of $ \mathfrak{M} $ are called {\em simplices}. For $ n \geq 0 $, a simplex $ S $ is 
called {\em $n$-dimensional} or {\em $n$-simplex} if number $ |S| $ of its elements equals $ n +1 $.

Let $(A,\mathfrak{M})$ be a simplicial scheme. 
For the computing its {\em homology groups $H_n(A,\mathfrak{M})$}, we define an arbitrary 
total order relation on  $A$. 
 Consider the complex
$$
0 \leftarrow \ZZ\mathfrak{M}_0 \stackrel{d_1}\leftarrow \ZZ\mathfrak{M}_1 
\stackrel{d_2}\leftarrow \ZZ\mathfrak{M}_2 \leftarrow \cdots \leftarrow 
\ZZ\mathfrak{M}_{n-1} \stackrel{d_n}\leftarrow \ZZ\mathfrak{M}_n \leftarrow \cdots
$$
where $\mathfrak{M}_n= 
\{(a_0, a_1, \ldots, a_n) | a_0< a_1< \cdots < a_n ~\&~ \{a_0, a_1, \ldots, a_n\}\in \mathfrak{M}\}$.
Elements of $\mathfrak{M}_n$ are called {\em ordered $n$-simplices}. 
Here $\ZZ\mathfrak{M}_n$ denotes the free Abelian group generated 
by ordered $n$-simplices.
The differentials $d_n$ are defined on ordered $n$-simplices by the formula
$$
 d_n(a_0, a_1, \ldots, a_n)= \sum_{i=0}^n (-1)^i (a_0, \ldots, \widehat{a_i} , \ldots, a_n)
$$
where $\widehat{a_i}$ denotes the operation of removing the symbol $a_i$ from the tuple.
We will suppose that the sets of $n$-simplices are finite. In this case, the differentials  
$d_n$ can be specified using integer matrices.

Each column of the matrix for $d_n$ corresponds to a tuple $ (a_0, a_1, \ldots, a_n)\in \mathfrak{M}_n$. 
Each string corresponds to $(a_0, \ldots, a_{n-1})\in \mathfrak{M}_{n-1}$. 
For each column $ (a_0, a_1, \ldots, a_n) $ and
string $ (a_0, \ldots, \widehat {a_i}, \ldots, a_n) $, at their intersection, the entry equals $ (-1)^i $.
Other entries of the matrix equal $0$.
For calculating the homology groups, each matrix $d_n$ is reduced to the Smith normal form.
The homology groups $H_n=Ker(d_n)/Im(d_{n+1})$ of this complex is equal to 
$$
	\ZZ^{|\mathfrak{M}_n|-rank(d_n)-rank(d_{n+1})}\oplus \ZZ/\delta_1\ZZ\oplus \cdots \oplus \ZZ/\delta_r\ZZ
$$ 
where $r=rank(d_{n+1})$ and $\delta_1, \cdots, \delta_r$ is the non-zero diagonal entries 
of the Smith normal form for the matrix $d_{n+1}$. 

\subsection{Homology groups of labelled asynchronous systems}

Let $(\mA, \lambda, L)$ be a labelled asynchronous system.

Introduce homology groups of the labelled asynchronous systems.
 For this purpose, consider the simplicial scheme $ (\lambda^+E, \fM) $ whose vertices are the elements
$ \lambda (a) $, where $ a \in E $ are elements for which there are
  $ s, s'\in S (s_0) $ satisfying $ (s, a, s') \in \Tran $. Thus
 
$$
	\lambda^+E = \{\lambda(a) ~|~ (\exists s\in S(s_0)) s\cdot a \in S\}.
$$
Simplices are finite sets $\{\lambda(a_1), \ldots, \lambda(a_k)\}$, $k\geq 1$,
for which the following two conditions hold:
\begin{itemize}
\item $(a_i, a_j)\in I$, for all $1\leq i<j \leq k$;
\item there are $s\in S(s_0)$ for which $s\cdot a_1\cdots a_k \in S$.
\end{itemize}

\begin{remark}
\begin{enumerate}
\item For every  $(s, a_1, \ldots, a_k)\in Q_k(\mA)$, we include the set $\{\lambda(a_1), \ldots, \lambda(a_k)\}\}$
in $\fM$.
\item If the elements are duplicated in $\{\lambda(a_1), \ldots, \lambda(a_k)\}$, then we remove them.
For example $\{a, b, a, c, a, b\}=\{a,b,c\}$.
\end{enumerate}
\end{remark}

\begin{definition}
{\em Homology groups  $H_n(\mA, \lambda,L)$} of a labelled asynchronous system is the homology groups  
$H_n(\lambda^+E, \fM)$ of the constructed simplicial scheme.
\end{definition}

\begin{example}
Consider an asynchronous system $\mA=(S, s_0, E, I, \Tran)$ where 
$S=\{000, 001,010,011,100,101,110 \}$, $s_0= 000$, $E=\{a_1, a_2, a_3\}$,
$I=\{(a_1,a_2), (a_2,a_1), (a_1,a_3), (a_3,a_1), (a_2,a_3), (a_3,a_2)\}$. 
Transitions correspond to arrows of the diagram:
$$
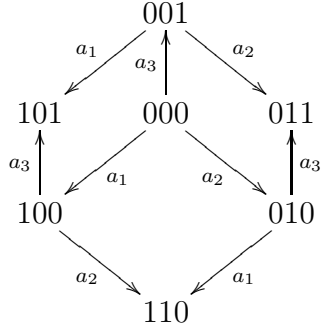
\xymatrix{
& 001 \ar[ld]_{a_1} \ar[rd]^{a_2} \\
101 & 000\ar[u]^{a_3} \ar[ld]^{a_1} \ar[rd]_{a_2} & 011\\
100 \ar[u]^{a_3} \ar[rd]_{a_2} & & 010 \ar[u]_{a_3} \ar[ld]^{a_1}\\
& 110
}
$$
Let $L=E$ and let the label function $\lambda: E\to L$ is defined as
 $\lambda(a)=a$ for all $a\in E$.
The simplicial scheme consists of vertices $E=\{a_1, a_2, a_3\}$ and simplices
$\{a_1, a_2\}$, $\{a_1, a_3\}$, $\{a_2, a_3\}$. 
Define the order on vertices by  
 $a_1<a_2<a_3$. 
Homology groups is computed by the complex
$$
0 \leftarrow \ZZ\{a_1, a_2, a_3\} \stackrel{d_1}\leftarrow \ZZ\{(a_1, a_2), (a_1, a_3), 
(a_2, a_3)\}
\leftarrow  0
$$
Matrix for $d_1$ equals
\begin{gather*}
\quad
\begin{array}{cccc}
 & ~~~~ (a_1,a_2) & ~~~(a_1,a_3) & ~~(a_2, a_3)
\end{array}
\\
\begin{array}{l}
a_1\\
a_2\\
a_3
\end{array}
\quad
\left(
\begin{array}{ccc}
~~~-1 & ~~~~~~~-1 & ~~~~~~~0\\
~~~~ 1 & ~~~~~~~~0 & ~~~~~~-1\\
~~~~ 0 & ~~~~~~~~1 & ~~~~~~~1
\end{array}
\quad
\right)
\end{gather*}
The Smith normal form for $d_1$ equals 
\begin{gather*}
\quad
\begin{array}{cccc}
 & ~~~~ (a_1,a_2) & ~~~(a_1,a_3) & ~~(a_2, a_3)
\end{array}
\\
\begin{array}{l}
a_1\\
a_2\\
a_3
\end{array}
\quad
\left(
\begin{array}{ccc}
~~~~1 & ~~~~~~~~0 & ~~~~~~~0\\
~~~~ 0 & ~~~~~~~~1 & ~~~~~~~0\\
~~~~ 0 & ~~~~~~~~0 & ~~~~~~~0
\end{array}
\quad
\right)
\end{gather*}
It follows that $H_0(\mA,\lambda,L)=\ZZ^{3-0-2}\oplus \ZZ/1\ZZ\oplus \ZZ/1\ZZ\cong \ZZ$, 
$H_1(\mA,\lambda,L)=\ZZ^{3-2-0}\cong \ZZ$. Other homology groups equal $0$.

The complex for computing groups $H_n(\mA(s),\lambda,L)$ for $s=001$ has unique non-zero term 
 $\ZZ\{a_1,a_2\}$. It follows 
$$
H_n(\mA(s),\lambda,L)=
\left\{
\begin{array}{cl}
\ZZ\oplus\ZZ, & \mbox{ if } n=0,\\
 0, & \mbox{ if } n>0.\\
\end{array}
\right.
$$

The complex for computing  $H_n(\mA(s),\lambda,L)$ for $s=011$ consists of zeros.
Therefore $H_n(\mA(011),\lambda,L)=0$ for all $n\geq 0$.
\end{example}

\begin{theorem}\label{isohomol}
If labelled asynchronous systems  $(\mA,\lambda, L)$ and $(\mA', \lambda', L)$ are
$Pom_L$-bisimilar, then their homology groups are isomorphic.
\end{theorem}
{\sc Proof.}
Denote by $\mathfrak{M}$ and $\mathfrak {M'}$ the simplicial schemes 
corresponded to the labelled asynchronous systems.
If the labelled asynchronous systems are $Pom_L$-bisimilar, then there is 
a labelled asynchronous system  together with the morphisms
$$
	(\mA, \lambda, L) \stackrel{(\sigma, \eta)}\longleftarrow
(\mA'', \lambda'', L) \stackrel{(\sigma', \eta')}\longrightarrow 
(\mA', \lambda', L) ~.
$$
Let $P^f(L)$ be the set of all finite subsets of $L$.
Consider a maps $\lambda_n: Q_n(\mA) \to P^f(L)$ acting as $\lambda(s, a_1, \ldots, a_n)= 
\{\lambda(a_1), \ldots, \lambda(a_n)\}$.
The function $\lambda$ can have equal values. Hence, the set
$\{\lambda(a_1), \ldots, \lambda(a_n)\}$ can contain $<n$ 
elements
 
 For $n=0$, we let $\lambda_0(s)= \emptyset$. By Proposition \ref{opensur} 
the maps  $Q_n(\sigma, \eta)$ and $Q_n(\sigma', \eta')$ are surjective.
The pairs $(\sigma,\eta)$ and $(\sigma',\eta')$ are morphisms of asynchronous 
systems. Hence, the following diagram is commutative
$$
\xymatrix{
Q_n(\mA) \ar[rd]_{\lambda_n} & Q_n(\mA'')\ar[d]^{\lambda_n''} \ar[l]_{Q_n(\sigma,\eta)} \ar[r]^{Q_n(\sigma',\eta')} & 
Q_n(\mA') \ar[ld]^{\lambda'_n}\\
 & P^f(L)
}
$$
We have the equalities
$
Im(\lambda_n )=Im(\lambda''_n)=Im(\lambda'_n ).
$
Consequently the simplicial sets  $\mathfrak{M}$ and $\mathfrak{M'}$ are equal. Therefore, the groups
$H_n(\mathfrak{M})$ and $H_n(\mathfrak{M'})$ are isomorphic.
\hfill $\Box$

\begin{corollary}\label{isoall}
Let $(\mA, \lambda, L)$ and $(\mA', \lambda', L)$ be $Pom_L$-bisimilar asynchronous systems.
For each $w= a_1\cdots a_k\in E^*$, $k\geqslant 0$, satifying  
$s_0\cdot w\in S$ there is a word $w'= a'_1\cdots a'_k\in E'^*$ such that
$s'_0\cdot w'\in S'$ and  
\begin{equation}\label{iso0} 
(\forall n\geqslant 0)~
H_n (\mA(s_0\cdot w), \lambda, L) \cong H_n(\mA'(s'_0\cdot w'), \lambda', L).
\end{equation}
\end{corollary}
{\sc Proof.} 
By Proposition  \ref{bisall}, in this case for the word $w$, there exists $w'$ 
for which  $(\mA(s_0\cdot w), \lambda, L)$ and  
$(\mA'(s'_0\cdot w'), \lambda', L)$ are $Pom_L$-bisimilar. 
Application of Theorem \ref{isohomol} to the obtained labelled 
asynchronous systems leads us to desired isomorphism of the homology groups.
\hfill $\Box$

\begin{example}
Consider well known labelled asynchronous systems
$$
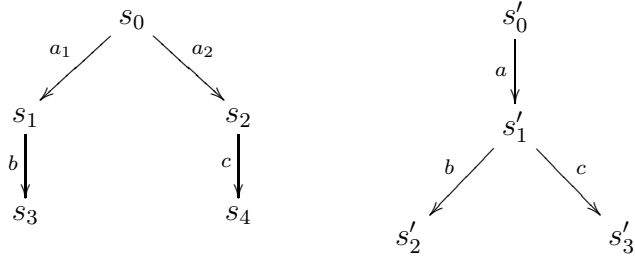
\xymatrix{
& s_0 \ar[ld]_{a_1} \ar[rd]^{a_2}\\
s_1 \ar[d]_b & & s_2\ar[d]_c\\
s_3 & & s_4
}
\qquad \qquad
\xymatrix{
& s_0' \ar[d]_{a}\\
& s_1' \ar[ld]_b \ar[rd]^c\\
s_2' & & s_3'
}
$$


The first asynchronous system  $\mA$ consists of 
 $S=\{s_0, s_1, s_2, s_3, s_4\}$,  $E=\{a_1, a_2, b, c\}$, 
 $I=\emptyset$, 
$\Tran = \{(s_0, a_1, s_1), (s_0, a_2, s_2), (s_1, b, s_3), (s_2, c, s_4)\}$.

The second asynchronous system $\mA'$ consists of 
 $S'=\{s'_0, s'_1, s'_2, s'_3\}$,  $E'=\{a, b, c\}$, 
 $I'=\emptyset$,  
$\Tran = \{(s'_0, a, s'_1),  (s'_1, b, s'_2), (s'_1, c, s'_3)\}$.

The label functions have values in 
 $L=\{a, b, c\}$ and are defined by
$$
\lambda(a_1)= \lambda(a_2)= \lambda'(a)= a,~ \lambda(b)= \lambda'(b)= b,~ 
\lambda(c)= \lambda'(c)= c.
$$

Compute $H_n(\mA(s_1), \lambda, L)$ by the complex 
$0\leftarrow \ZZ\{b\}\leftarrow 0$.
We have 
$$
H_n(\mA(s_1), \lambda, L)=
\left\{
\begin{array}{cl}
\ZZ, & \mbox{ if } n=0,\\
 0, & \mbox{ if } n>0.\\
\end{array}
\right.
$$

The groups  $H_n(\mA'(s_1'), \lambda', L)$ are isomorphic to homology groups of the complex 
$0\leftarrow \ZZ\{b\}\oplus \ZZ\{c\}\leftarrow 0$. We have 
$$
H_n(\mA'(s_1'), \lambda', L)=
\left\{
\begin{array}{cl}
\ZZ\oplus\ZZ, & \mbox{ if } n=0,\\
 0, & \mbox{ if } n>0.\\
\end{array}
\right.
$$
The groups $H_0(\mA(s_1), \lambda, L)$ and  $H_0(\mA'(s_1'), \lambda', L)$ are not isomorphic.
It follows from Corollary \ref{isoall} that $(\mA, \lambda, L)$ and $(\mA', \lambda', L)$ are not 
$Pom_L$-bisimilar.
\end{example}

\section{Homology groups of labelled Petri nets}

Recall some definitions from theory of Petri nets.
Then consider homology groups of labelled Petri nets and prove that 
for each simplicial scheme, there is a labelled Petri net homological equivalent to this simplicial 
scheme.

\subsection{Petri nets}

We view ``display'' and ``function'' as synonyms.
For a finite set $P$, let $\NN^P$ denotes a set of all functions 
$M: P\to \NN$, 
where 
$\NN= \{0, 1, 2, \ldots\}$ is the set of non-neganbve integers.
For any $M_1, M_2\in \NN^P$, define a sum $M_1+M_2$ as a function with values
$(M_1+M_2)(p)= M_1(p)+ M_2(p)$ for all $p\in P$. 
Let $M_1 \geq M_2$ if $M_1(p)\geq M_2(p)$ for all $p\in P$.
If $M_1\geq M_2$, then we can define a {\em difference $M_1- M_2$} 
as the function with the values 
$M_1(p)- M_2(p)$.
Define a {\em scalar product} by $M_1\cdot M_2= \sum_{p\in P} M_1(p)M_2(p)$. 

A {\em Petri net} $\mN= (P, T, pre, post, M_0)$ consists of finite sets 
$P$ and $T$ with  two maps  $pre: T\to \NN^P$, $post: T\to \NN^P$ and a function $M_0: P\to \NN$
 called {\em initial marking}. Elements $p\in P$ are called {\em places}, and 
$t\in T$ are {\em events}. 
A {\em marking} is an arbitrary function $M: P\to \NN$. 

\begin{figure}[ht]
\begin{center}
\begin{picture}(100,70)
\put(15,40){\line(1,0){15}}\put(30,40){\line(0,1){12}}
\put(30,48){\vector(1,0){20}}
\put(30,42){\vector(1,0){20}}
\put(30,41){\vector(1,-1){20}}
\put(18,43){$t_1$}
\put(15,40){\line(0,1){12}}\put(15,52){\line(1,0){15}}
\put(85,40){\line(1,0){15}}\put(100,40){\line(0,1){12}}
\put(88,43){$t_2$}
\put(85,40){\line(0,1){12}}\put(85,52){\line(1,0){15}}
\put(85,10){\line(1,0){15}}\put(100,10){\line(0,1){12}}
\put(85,10){\line(0,1){12}}\put(85,22){\line(1,0){15}}
\put(88,13){$t_3$}
\put(57,45){\circle{14}}
\put(47,1){$p_2$}

\put(57,48){\circle*{3}}
\put(57,44){\circle*{3}}
\put(47,58){$p_1$}

\put(65,46){\vector(1,0){20}}

\put(57,15){\circle{14}}
\put(57,15){\circle*{3}}
\put(63,21){\vector(1,1){20}}
\put(65,15){\vector(1,0){20}}

\end{picture}
\end{center}
\caption{Example of Petri net} \label{netpic}

\end{figure}
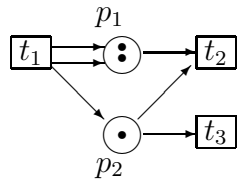

A Petri net can be given as a directed graph whose vertices are places 
depicted by circles, and events depicted by rectangles.
Every arrow goes from an event to a place or from a place to an event.
For any  $t\in T$, the number entering into it arrows equals $pre(t)(p)$ 
and the number of arrows outgoing from $t$ equals
$post(t)(p)$.
The initial marking is given by drawing the points in each place. These points are called {\em tokens}.
 The number of tokens in a place $p$ is equal to $M_0(p)$.
If $ M_0 (p) = 0 $, then the place is empty.

Fig. \ref{netpic} shows a Petri net $\mN= (P,T, pre, post, M_0)$
where $P=\{p_1, p_2\}$, $T=\{t_1, t_2, t_3\}$. The values $pre(t_i)(p_j)$ and $post(t_i)(p_j)$, 
$1\leq i\leq 3$, $1\leq j\leq 2$, are equal to the entries of the matrices
$$
(pre(t_i)(p_j)) = \left(
\begin{array}{cc}
0 & 0\\
1 & 1\\
0 & 1
\end{array}
\right)
\qquad
(post(t_i)(p_j)) = \left(
\begin{array}{cc}
2 & 1\\
0 & 0\\
0 & 0
\end{array}
\right)
$$

\subsection{Labelled asynchronous system for a Petri net and its homology groups}

Let $\mN=(P,T, pre, post, M_0)$ be a Petri net.
Consider a corresponding asynchronous system $\mA(\mN)= (S, s_0, E, I, \Tran)$, 
with $S=\NN^P$, $s_0=M_0$, $E = T$.
The relation of independence $I$ consists of pairs $(e_1, e_2)\in T\times T$ for which
the scalar product  $(pre(e_1)+post(e_1)\cdot (pre(e_2)+post(e_2))$ equals $0$.
This means that $ e_1 $ and $ e_2 $ do not have common input or output places.
The set $\Tran$ consists of triples $(M, e, M')$ where $M$ and $M'$ are markings  
 and $e\in T$ satifies two following conditions
\begin{itemize}
\item $M\geq pre(e)$,
\item $M-pre(e)+post(e)=M'$.
\end{itemize} 

\begin{figure}[ht]
\begin{center}
\begin{picture}(100,70)
\put(15,40){\line(1,0){15}}\put(30,40){\line(0,1){12}}
\put(30,48){\vector(1,0){20}}
\put(30,42){\vector(1,0){20}}
\put(30,41){\vector(1,-1){20}}
\put(18,43){$t_1$}
\put(15,40){\line(0,1){12}}\put(15,52){\line(1,0){15}}
\put(85,40){\line(1,0){15}}\put(100,40){\line(0,1){12}}
\put(88,43){$t_2$}
\put(85,40){\line(0,1){12}}\put(85,52){\line(1,0){15}}
\put(85,10){\line(1,0){15}}\put(100,10){\line(0,1){12}}
\put(85,10){\line(0,1){12}}\put(85,22){\line(1,0){15}}
\put(88,13){$t_3$}
\put(57,45){\circle{14}}
\put(47,1){$p_2$}

\put(57,45){\circle*{3}}
\put(47,58){$p_1$}

\put(65,46){\vector(1,0){20}}

\put(57,15){\circle{14}}
\put(63,21){\vector(1,1){20}}
\put(65,15){\vector(1,0){20}}
\end{picture}
\end{center}
\caption{The marking obtained by operation of the event $t_2$} \label{newmark}

\end{figure}
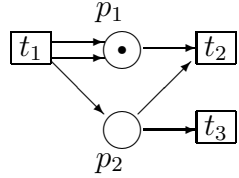

If $(M,e,M')\in \Tran$, then we say that the marking $M'$ is obtained from $M$ {\em by operation of event}
 $e\in T$. For example, for Petri net in Fig. \ref{netpic}, 
we have $pre(t_2)\geq M_0$. 
The operation of the event  $t_2$ leads  
to the new magking  $M_1= M_0- pre(t_2)+post(t_2)$ (Fig. \ref{newmark}).

Let $L$ be an arbitrary nonempty set. A Petri net $\mN$
with a function $\lambda: T \to L$ is called  {\em labelled}. 
The asynchronous system $\mA(\mN)$ corresponding $\mA$ has 
the set of events $E=T$. 
Hence, for any labelled Petri nets, it is defined the labelled asynchronous system  
 $(\mA(\mN), \lambda, L)$.

\begin{definition}
Let $(\mN, \lambda, L)$ be a labelled Petri net. Its homology groups $H_n(\mN,\lambda,L)$
are defined as $H_n(\mA(\mN), \lambda, L)$, $n\geq 0$.
\end{definition}

\begin{example}
Consider the Petri net $\mN=(P,T, pre, post, M_0)$, in Fig. \ref{exhomnet}.
Let $L=E=\{t_1, t_2, t_3, t_4\}$, $\lambda(t_i)=i$, for all $1\leq i\leq 4$.
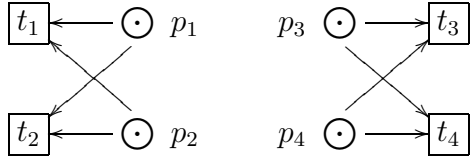
\begin{figure}[ht]
$$
\xymatrix{
 *+[F]{t_1} & \bigodot \ar[l]\ar[ld]~p_1 & p_3~ \bigodot \ar[r]\ar[rd] & *+[F]{t_3}\\
 *+[F]{t_2} & \bigodot \ar[l]\ar[lu]~p_2 & p_4~ \bigodot \ar[r]\ar[ru] & *+[F]{t_4}\\
}
$$
\caption{Example of computing the homology groups of Petri net}\label{exhomnet}
\end{figure}
The relation  $I$ contains the pairs $(t_1,t_3)$, $(t_1, t_4)$, $(t_2, t_3)$, $(t_2, t_4)\}$, 
$(t_3, t_1)$, $(t_3, t_2)$, $(t_4, t_1)$, $(t_4,t_2)$.
The simplicial set $(E,\fM)$ give the following sets of simplices
$$
\fM_0 = \{t_1, t_2, t_3, t_4\},\\
\fM_1 = \{(t_1,t_3), (t_1, t_4), (t_2, t_3), (t_2, t_4)\},~
$$ 
and $\fM_n = \emptyset$ for $n\geq 2$. We get the following complex for the computing 
the homology groups of the labelled Petri nets:
$$
0 \leftarrow \ZZ^4 \stackrel{d_1}\longleftarrow \ZZ^4 \leftarrow 0.
$$
The differential $d_1$ is given by the matrix
\begin{gather*}
\quad
\begin{array}{cccccc}
 & ~~(t_1,t_3) & (t_1,t_4) & (t_2,t_3) & (t_2,t_4)
\end{array}
\\
\begin{array}{l}
t_1\\
t_2\\
t_3\\
t_4
\end{array}
\left(
\begin{array}{cccc}
-1 & ~~~~~ -1 & ~~~~~~0 & ~~~~~~0\\
~0 & ~~~~~~ 0 & ~~~~~ -1 & ~~~~~ -1\\
+1 & ~~~~~~ 0 & ~~~~~ +1 & ~~~~~~0\\
~0 & ~~~~~ +1 & ~~~~~~0 & ~~~~~ +1
\end{array}
\right)
\end{gather*}
Its Smith normal form has the diagonal entries $(1, 1, 1, 0)$.
Consequently 
$$
H_0(\mN, \lambda, L)\cong H_1(\mN, \lambda, L)=\ZZ \mbox{ and } H_n(\mN, \lambda, L)=0
\mbox{ for all } n\geq 2.
$$
\end{example}

A sequence of Abelian groups $A_k$, $k\geq 0$, is called to be {\em finite} if there is $n\geq 0$
such that $A_k=0$ for all $k>n$.
\begin{theorem}\label{homall}
For an arbitrary finite sequence of finitely generated Abelian groups $A_0$, $A_1$, $A_2$, \ldots where 
  $A_0$ is free and is not equal to $0$,
there exists a labelled Petri net such that its $k$th homology groups are isomorphic to $A_k$ 
for all $k\geq 0$.
\end{theorem}
{\sc Proof.} In this case by \cite[Chapter 4, Exercise C-7]{spa1966}, there exists a compact polyhedron
with homology groups  $A_k$ for all $k\geq 0$. 
Compact polyhedra
are precisely the topological spaces admitting
triangulations \cite[Chapter 3, Corollary 20]{spa1966}.
Hence, there exists a simplicial scheme  $(X, \fM)$ the homology groups of which 
are isomorphic to $A_k$.

 Let $(E,\fM')$ be a {\em barycentric subdivision} of the simplicial set $(E, \fM)$. 
Vertices $e\in E$ of the barycentric subdivision are simplices $ \sigma \in \fM $.
Simplices of $(E,\fM')$ are finite sets of simplices
$\{\sigma_0, \ldots, \sigma_n\}$ totally ordered by the relation $\subseteq$.
It means that there is a permutation $(\sigma_{i_0}, \ldots, \sigma_{i_n})$ such that
$\sigma_{i_0}\subset \sigma_{i_1}\subset \ldots \subset \sigma_{i_n}$.
It is well known that homology groups of $(E, \fM')$ are isomorphic to
homology groups of $(X,\fM)$.
Define a relation $I$ on $E$ by 
$$
(\sigma,\sigma')\in I \Leftrightarrow \sigma\subset \sigma' \vee \sigma'\subset \sigma.
$$

\unitlength=0.7pt
\begin{figure}[ht]
\begin{center}
\begin{picture}(400,140)

\put(50,80){\circle{20}}
\put(50,80){\circle*{3}}

\put(150,80){\circle{20}}
\put(150,80){\circle*{3}}

\put(350,80){\circle{20}}
\put(350,80){\circle*{3}}

\put(225,55){\circle{3}}
\put(250,55){\circle{3}}
\put(275,55){\circle{3}}

\put(67,80){$p_1$}
\put(167,80){$p_2$}
\put(367,80){$p_m$}
\put(65,28){$e_1$}
\put(165,28){$e_2$}
\put(365,28){$e_m$}

\put(40,20){\line(1,0){20}}
\put(40,20){\line(0,1){20}}
\put(60,40){\line(-1,0){20}}
\put(60,40){\line(0,-1){20}}
\put(50,70){\vector(0,-1){30}}

\put(140,20){\line(1,0){20}}
\put(140,20){\line(0,1){20}}
\put(160,40){\line(-1,0){20}}
\put(160,40){\line(0,-1){20}}
\put(150,70){\vector(0,-1){30}}

\put(340,20){\line(1,0){20}}
\put(340,20){\line(0,1){20}}
\put(360,40){\line(-1,0){20}}
\put(360,40){\line(0,-1){20}}
\put(350,70){\vector(0,-1){30}}

\end{picture}
\end{center}
\caption{The constructing of a Petri net}\label{step1}
\end{figure}
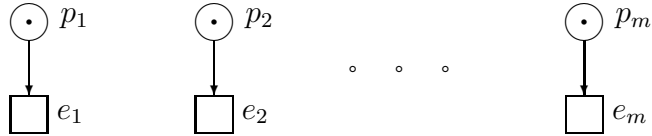

Building a Petri net is similar to the construction of the work \cite{X20122}. 
Denote the elements of $E$ by $e_1$, $e_2$, \ldots, $e_m$ where $m=|E|$.
Consider the Petri net depicted in Fig. \ref{step1}. It consists of places $p_i$, 
 connected with the events $e_i$ by the arrows where $i= 1, 2, \ldots, m$.  
The initial marking is defined as $M_0(p_i)=1$ for all $i=1, 2, \ldots, e_m$.
For every $(e_i, e_j)\notin I$, 
we make the events $e_i$ and $e_j$ to be dependent by
adding two arrows as shown in Fig. \ref{step2}.

\unitlength=0.7pt
\begin{figure}[ht]
\begin{center}
\begin{picture}(200,100)

\put(50,70){\circle*{3}}
\put(50,70){\circle{20}}
\put(150,70){\circle*{3}}
\put(150,70){\circle{20}}

\put(27,70){$p_i$}
\put(162,70){$p_j$}

\put(45,18){$e_i$}
\put(145,18){$e_j$}

\put(40,10){\line(1,0){20}}
\put(40,10){\line(0,1){20}}
\put(60,30){\line(-1,0){20}}
\put(60,30){\line(0,-1){20}}
\put(50,60){\vector(0,-1){30}}

\put(140,10){\line(1,0){20}}
\put(140,10){\line(0,1){20}}
\put(160,30){\line(-1,0){20}}
\put(160,30){\line(0,-1){20}}
\put(150,60){\vector(0,-1){30}}
\put(150,60){\vector(-3,-1){92}}

\put(50,60){\vector(3,-1){92}}

\end{picture}
\end{center}
\caption{Adding arrows to the Petri net}\label{step2}
\end{figure}
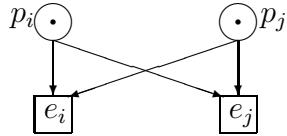

Let $L=E$ and let the label function defined as $\lambda(e_i)=e_i$ for all $i=1, \ldots, m$.
For every $e_i\in E$, we have $s_0\cdot e_i\in S$. It follows that the set 
of vertices of a simplicial scheme corresponding to the Petri net is equal to
 $E$.
For each nonempty subset $\{e_{i_0}, \ldots, e_{i_n}\} \subseteq E$ consisting  
of mutually independent elements, we have $s_0\cdot e_{i_0}\cdots e_{i_n} \in S$.
Consequently the simplicial set corresponding to the Petri net is equal to  $(E, \fM')$. 
Thus, $H_n(\mN, \lambda, L)=A_n$ for all $n\geq 0$.
\hfill $\Box$

\begin{corollary}
For any finite sequence of finitely generated Abelian groups $A_0$, $A_1$, $A_2$, \ldots where 
  $A_0$ is free and non-zero,
there is a labelled asynchronous system the $k$th homology groups of which are isomorphic to $A_k$ 
for all $k\geq 0$.
\end{corollary}

\end{document}